\documentclass[twocolumn]{article}
\usepackage[utf8]{inputenc}
\usepackage{authblk}
\usepackage{setspace}
\usepackage[margin=1in]{geometry}
\usepackage{graphicx}
\graphicspath{ {./Images/} }

\usepackage{subcaption}
\usepackage{amsmath}
\usepackage{multicol}
\usepackage{multirow}
\usepackage{xcolor}
\usepackage{tabularx}
\usepackage{amssymb}
\usepackage{booktabs}
\usepackage{float}
\usepackage{dblfloatfix}
\usepackage{enumitem}
\usepackage{titlesec}
\usepackage{hyperref}
\usepackage{xurl}
\usepackage{verbatim}
\usepackage{colortbl}
\usepackage{threeparttable}
\titleformat{\paragraph}[block]{\normalfont\normalsize\bfseries}{\theparagraph}{1em}{}

\usepackage[
    backend=biber,
    style=numeric-comp,
    sorting=none,
    maxbibnames=10,
    minbibnames=10
]{biblatex}

\AtBeginBibliography{\footnotesize}

\DeclareFieldFormat*{url}{\url{#1}}
\renewbibmacro*{doi+eprint+url}{%
  \printfield{doi}%
  \newunit\newblock
  \iftoggle{bbx:eprint}{\usebibmacro{eprint}}{}%
  \newunit\newblock
  \iffieldundef{doi}
    {\usebibmacro{url+urldate}}
    {}}

\renewbibmacro*{urldate}{
  \ifentrytype{online}
    {\iffieldundef{urlyear}{}{
      {[}Accessed on:
      \thefield{urlyear}
      \shortmonth{\thefield{urlmonth}}\addspace
      \thefield{urlday}{]}}}
    {}}

\title{Cultural multilevel selection of institutions for sustainable social metabolism}

\author[1*]{Tim Zandee}
\author[2]{Timothy M. Waring}
\author[3, 4]{Andrew K. Ringsmuth}

\affil[1]{Copernicus Institute of Sustainable Development, Utrecht University, Netherlands}
\affil[2]{School of Economics and Mitchell Center for Sustainability Solutions, University of Maine, USA}
\affil[3]{Wegener Center for Climate and Global Change, University of Graz, Austria}
\affil[4]{Complexity Science Hub, Vienna, Austria}
\affil[*]{Address correspondence to: tim.zandee@gmail.com}

\titlespacing*{\section}{0pt}{1ex plus .2ex minus .2ex}{1ex}      
\titlespacing*{\subsection}{0pt}{1ex plus .2ex minus .2ex}{0.5ex} 
\titlespacing*{\paragraph}{0pt}{1ex plus .2ex minus .2ex}{1em}    

\begin{document}

\maketitle

\begin{abstract}
Like all living systems, human societies persist by acquiring energy and materials, coordinating internally and adapting to their environments. Sustainability requires maintaining this social metabolism within ecological limits. Sociometabolic research quantifies the energy and material flows sustaining societies, commons scholarship identifies principles by which communities sustain shared resources, and cultural multilevel selection explains how such institutions spread through intergroup competition. Yet these perspectives developed separately, leaving untested whether such institutions can evolve under physical metabolic constraints. Here we develop a stock-flow-consistent agent-based model in which metabolically constrained consumers compete for a renewable resource and culturally inherit harvesting strategies, group markers and institutional rules embodying three of Ostrom's design principles: user boundaries, resource boundaries and monitoring. Decomposing selection into individual and group components shows that the outcome hinges on the social dilemma's level relative to that of selection. With one global resource, both individual and group selection favour unrestrained harvesting; resources and populations always collapse. With local resources, group selection for restraint overcomes individual selection against it, restrained harvesters persist and institutions matching Ostrom's principles dominate through group selection. Sustainable groups can evolve under individual selection for over-consumption, even under physical metabolic constraints, when the dilemma lies below the level of selection.
\end{abstract}


\section{Introduction}
\subsection{Life's multifaceted persistence problem}\label{sec:intro:persistence}
Sustainable development entails maintaining a social foundation of human well-being within ecological limits \cite{raworth_doughnut_2017}. It is also a particular instance of a more general problem: how systems that maintain themselves far from thermodynamic equilibrium persist through time. Such systems, from cells to societies, must acquire low-entropy energy and materials, use them to operate, maintain and develop their internal organisation, and dissipate high-entropy wastes. These requirements are mutually constitutive: metabolism cannot occur without coordination among a system's components, coordinated organisation cannot persist without metabolic provisioning, and neither can it persist without an environment that supplies usable resources and assimilates wastes. Moreover, no system can maintain itself indefinitely, so long-term persistence depends further on transmitting the organisational information needed to construct new self-maintaining systems, and on evolutionary processes that keep those systems fitted to changing environments. Environmental coupling, internal coordination, metabolism and evolution are thus complementary dimensions of one persistence problem, operating across levels of organisation and spatio-temporal scales.

Human societies are no exception. Like life at smaller scales, they operate, maintain themselves and grow through continuous exchanges of energy and materials with their environment, a process described as \emph{social metabolism} \cite{haberl_contributions_2019}. Despite their distinctive intangible dimensions, such as trust and value, social systems rest on material and energetic processes that remain subject to the physical constraints governing all living systems. The human capacity to exploit environmental resources enabled our species to occupy diverse ecosystems and supported unprecedented growth in population, technological complexity and wealth; groups able to coordinate the mobilisation of greater material and energetic resources have tended to outcompete others, creating a positive feedback between social expansion and environmental exploitation \cite{waring_characteristic_2023}. Over longer timescales, however, such societies increasingly alter the ecological conditions on which their own metabolism depends, threatening their persistence \cite{richardson_earth_2023, fanning_doughnut_2025}.

Sustaining social metabolism is also a coordination problem, studied extensively in commons scholarship as a collective action problem arising from social dilemmas \cite{ostrom_collective_2000}: maintaining a society requires aligning the actions of many individuals whose immediate interests diverge from those of the collective, and hence institutions that foster cooperation among resource users. Ostrom's design principles characterise institutions that have coordinated resource users to sustain shared ecological systems over long periods \cite{ostrom_collective_2000, cox_review_2010, wilson_core_2020, wilson_generalizing_2013}. Analogous coordination problems arise at every level of biological organisation, from microbial communities to multicellular organisms and their groups, and evolutionary theory has long sought to explain how they are resolved \cite{west_major_2015}. Several authors have consequently argued that evolutionary theory provides a promising foundation for sustainability science \cite{kline_designing_2018, ostrom_institutions_2014, safarzynska_evolutionary-economic_2013, waring_multilevel_2015, wilson_multilevel_2023, ostrom_collective_2000}.

These literatures thus address complementary dimensions of life's persistence problem at the level of human societies, and each supplies essential components for modelling how societies can solve it, which we now briefly review: \S\ref{sec:intro:metabolism} examines the metabolic constraints that societies share with other living systems, and \S\ref{sec:intro:selection} the multilevel selection processes through which cooperation over shared resources evolves.

\subsection{Multilevel metabolism}\label{sec:intro:metabolism}
In biology and ecology, the use of energy and materials scales with system size in remarkably consistent ways. Basal metabolic rate increases with body mass $M$ as a power law $aM^{b}$, with $a$ a normalisation constant and $b$ the scaling exponent; across a broad range of organisms $b \approx 3/4$, a relationship known as Kleiber's law \cite{brown_toward_2004, agutter_metabolic_2004}. Because the power available for acquiring resources is itself metabolically constrained, maximum rates of resource acquisition are conventionally taken to scale in the same way \cite{brown_toward_2004, sibly_representing_2013}. Since $b<1$, larger organisms need more energy in absolute terms but less per unit mass, and this size-dependent structure of energy budgets propagates into growth, reproduction and population dynamics \cite{sibly_representing_2013}.

However, the generality of 3/4-power scaling is contested. Similar scaling has been reported at higher levels of biological organisation, including whole ecosystems \cite{fant_biomass_2024}, but the exponent shifts across major evolutionary transitions, from superlinear in prokaryotes through approximately linear in protists to sublinear in metazoans \cite{delong_shifts_2010}, and varies systematically with taxon, physiological state, activity level, environment and size range \cite{glazier_beyond_2005, glazier_unifying_2010, isaac_why_2010, glazier_variable_2022}. Theoretical explanations, whether from the geometry of resource-distribution networks or from surface-area and volume constraints on metabolic demand, remain controversial, and none of them accommodates the full observed range \cite{west_general_1997, west_fourth_1999, glazier_unifying_2010, agutter_metabolic_2004, savage_sizing_2008}. 3/4-power scaling nonetheless remains a well-supported central tendency for metazoans \cite{delong_shifts_2010, isaac_why_2010}, and we adopt it as an idealisation of an empirical regularity rather than a physical constant.

Analogous scaling appears in social groups across species, including human societies \cite{brown_energetic_2011, zhang_allometric_2010}. Whole colonies of social insects consume energy at rates that scale sublinearly with colony mass, with a mean exponent near $0.76$ \cite{hou_energetic_2010, waters_differentiating_2017}. This is a property of the functioning colony rather than aggregation over its individuals: it disappears when the same individuals are measured as unstructured groups \cite{waters_allometric_2010}. How a group is organised therefore shapes how its collective metabolism scales, making group-level metabolic patterns outcomes of social organisation and not physics alone.

Human societies show regularities of the same kind, with exponents that differ by variable and level of organisation. Settled area scales sublinearly with population in both pre-Columbian and contemporary societies, while urban socioeconomic output scales superlinearly \cite{bettencourt_growth_2007, ortman_pre-history_2014}, and estimates are sensitive to how social units are bounded \cite{arcaute_constructing_2015}. In the most direct parallel to biological metabolic scaling, \textcite{leiva_rules_2022} report that humanity's energy consumption scales with its accumulated material stocks with $b \approx 0.78$, implying that growing energy throughput entails increasing material requirements, and hence environmental pressures, irrespective of the energy source.

Remarkably, despite these regularities, human societies have partially escaped the energetic constraints governing other species through the expansion of exosomatic metabolism. Technology and fossil energy have dramatically expanded energy throughput per capita and per unit area, so modern societies consume energy far beyond physiological requirements to support infrastructure, transport, production and other material provisioning \cite{burger_extra-metabolic_2017}. Consistent with this, energy use in industrialized nations has scaled superlinearly with population, though variably over time \cite{delong_socio-economic_2015}. This expansion has enabled unprecedented socioeconomic development while increasing humanity's demands on a finite biosphere \cite{burger_extra-metabolic_2017}. Yet the persistence of near-biological scaling between energy use and material stocks indicates that biophysical constraints are not dissolved by industrialization but remain detectable in contemporary societies \cite{leiva_rules_2022}, bearing on the limits to economic growth \cite{brown_energetic_2011, murphy_limits_2022}.

At any level of organisation, such scaling patterns are aggregate outcomes rather than mechanisms: they describe how energy use covaries with collective scale, not how the organisation of resource use, and in human societies the institutions governing it, produce or respond to those patterns. Understanding how these patterns arise, and how far they can be steered, requires mechanisms relating individual persistence to group persistence. These are the subject of multilevel evolutionary theory.

\subsection{Multilevel selection of sustainable institutions}\label{sec:intro:selection}
\begin{figure*}[b]
    \centering
    \includegraphics[width=1\linewidth]{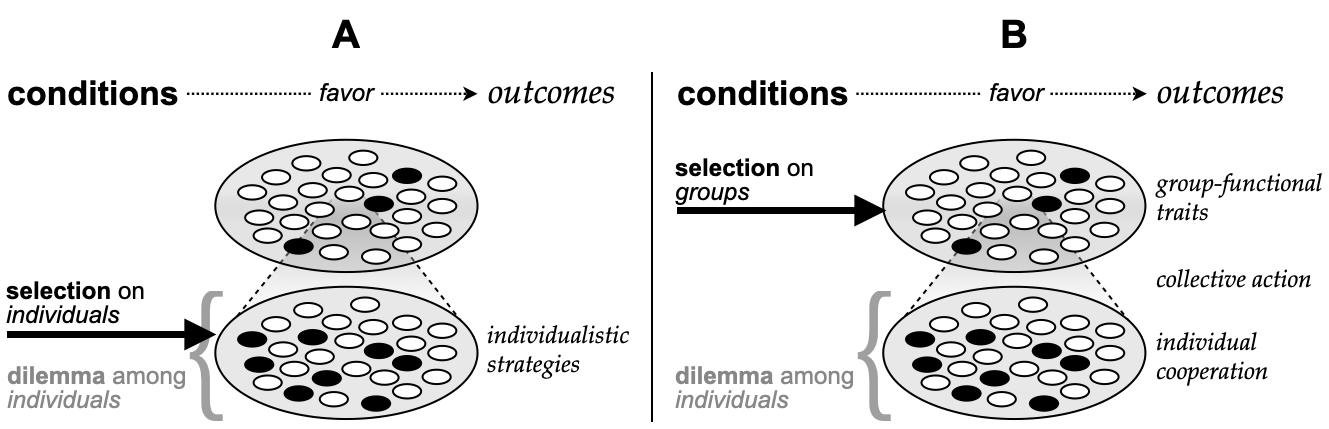}
    \caption{The cultural multilevel selection (CMLS) framework predicts that group-level cultural traits for resource management (e.g. institutions and cooperation) will evolve when the dominant level of cultural selection is above that of the dilemma over resource use
    \protect\cite{waring_multilevel_2015}. }
    \label{fig:CMLS Framework}
\end{figure*}

Resource dilemmas occur at all scales of life. In microbes, thermodynamics imposes a trade-off between the rate and yield of ATP production \cite{pfeiffer_cooperation_2001, maclean_tragedy_2008}, which creates a social dilemma over shared resources: individual reproductive rate is maximised by consuming as fast as possible, but average fitness is highest if all individuals use resources efficiently, making efficient use a primitive form of cooperation \cite{pfeiffer_cooperation_2001, maclean_tragedy_2008}. Resolving such dilemmas at any scale requires resource users to cooperatively restrain over-extraction, raising the question of how cooperation can evolve when selection rewards individuals who over-extract. Multilevel selection (MLS) theory states that while selection favours selfish behaviour within groups, competition between groups can favour those whose members cooperate more \cite{west_sixteen_2011, wilson_core_2020}, making cooperation a collective adaptation rather than an anomaly. In well-mixed populations, the efficient-use strategy can proliferate only where there is no dilemma \cite{amado_competing_2016}. In a model with explicit groups that grow and divide, selection acts at both levels and efficient use is favoured over a far broader range of conditions, most strongly when groups are small and migration between them rare \cite{amado_competing_2016}.

The same mechanism is argued to apply to human cultures \cite{wilson_generalizing_2013, wilson_multilevel_2023}, where adaptation proceeds through both genetic and cultural inheritance \cite{wilson_evolution_2013}. Cultural multilevel selection (CMLS) extends MLS to socially transmitted beliefs, norms, technologies and institutions, which vary, are selected and transmitted at both the individual and group levels \cite{richerson_cultural_2016, wilson_multilevel_2023, smaldino_cultural_2014}, and which evolve more rapidly than genes \cite{perreault_pace_2012}. It has been argued that cultural group selection is essential for explaining human cooperation \cite{richerson_cultural_2016} and the evolution of complex societies \cite{turchin_warfare_2010}. Recent work also suggests that it may be shifting the dominant level of selection in human evolution from genetic individuals to cultural groups \cite{waring_long-term_2021, waring_cultural_2025}, a process proposed to have caused the Anthropocene while also obstructing its global solutions \cite{waring_characteristic_2023}.

CMLS thus predicts that institutions enhancing group persistence spread through competition between groups, and can explain the emergence of sustainable resource-management institutions such as those documented by Ostrom. Ostrom herself proposed that they emerge through cultural evolution \cite{ostrom_institutions_2014}, anticipating arguments that cultural group selection favours such institutions \cite{waring_multilevel_2015}, and models in which they emerge endogenously even while individuals remain under selective pressure to overexploit shared resources \cite{waring_coevolution_2017}. In the framework of \textcite{waring_multilevel_2015}, the outcome of an environmental dilemma depends on how the dominant level of selection compares with the social scale of the dilemma (figure \ref{fig:CMLS Framework}). When the dominant level of selection lies below that of the dilemma, for example on individuals within groups, selection favours individualistic strategies and the dilemma goes unresolved. When it lies at or above that level, for example on groups themselves, selection favours group-functional traits and a resolution.

\subsection{Linking metabolism, coordination and evolution}\label{sec:intro:linking}
Although CMLS studies, commons scholarship and sociometabolic research address complementary aspects of societies' persistence problem, they have developed largely independently. Cultural-evolutionary models of institutions, including CMLS models, typically represent socioeconomic interactions while abstracting away a society's biophysical basis \cite{ghorbani_managing_2016, andrews_cultural_2024, hebert-dufresne_source-sink_2022}. Commons scholarship has identified robust principles for governing shared resources, but tends to represent biophysical processes in abstract terms without applying physical principles, and has only recently begun to embed them in explicitly evolutionary frameworks \cite{wilson_generalizing_2013, waring_multilevel_2015, kline_designing_2018}. Sociometabolic research has analysed biophysical interactions between nature and societies in physical terms \cite{haberl_contributions_2019} while paying less attention to the cultural-evolutionary processes that transform patterns of resource use. This separation is striking because metabolism, coordination and adaptation are inseparable dimensions of a society's persistence.

In this paper, we begin to integrate these perspectives in a multilevel selection model of competing resource consumers with energy budgets, endogenous group dynamics, cultural inheritance and varying institutional strategies. Following the recommendations of \textcite{sibly_representing_2013}, we incorporate (socio-)metabolic ecology into the CMLS framework of \textcite{waring_coevolution_2017}: consumers carry somatic energy budgets under explicit metabolic constraints, and group-level structure and its metabolic consequences emerge from their interactions rather than being imposed as a group-level scaling relation. We refine the framework of \textcite{waring_multilevel_2015} by modelling the resource dilemma at different levels and letting consumers vary in the social scope of the institutional rules they follow. That variation operationalises three of Ostrom's design principles, in the reformulation of \textcite{cox_review_2010}: user boundary (1A), resource boundary (1B) and monitoring (4). We ask how institutions regulating resource use evolve under physical metabolic constraints, and whether cultural group selection can favour forms of social organisation able to maintain both metabolically constrained societies and the ecological systems on which they depend.

\section{Materials and methods} \label{sec:Materials and methods}

\begin{figure*}[b]
    \centering
    \includegraphics[width=1\linewidth]{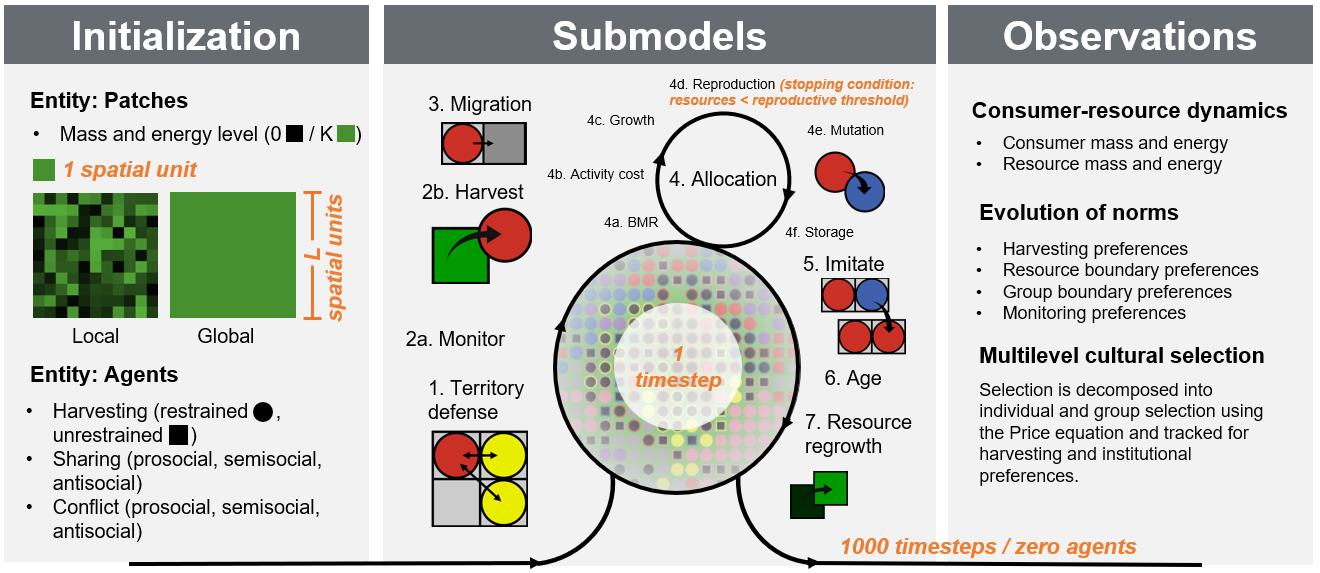}
    \caption{Model structure in vODD format \protect\cite{szangolies_visual_2024}.}
    \label{fig:vODD}
\end{figure*}

We created a model combining endogenous group selection dynamics among consumers (allowing for cooperation-defection dynamics) with physically constrained metabolic processes (ensuring physically plausible material and energy transformations). The model is built in NetLogo 7.0.4 and is run and analysed using LogoLink 1.0.0 and R 4.6.1 \cite{wilensky_netlogo_1999, vartanian_logolink_2026, r_core_team_r_2024}. This section provides a summary description, visualised in figure \ref{fig:vODD}. A complete model description using the Overview, Design concepts and Details (ODD) protocol for describing ABMs \cite{grimm_odd_2020, grimm_standard_2006}, is included in Appendix~\ref{Appendix A} and available together with the model and analysis code through \href{https://github.com/Tim-Zandee/The-cultural-evolution-of-sustainable-societies}{github.com/Tim-Zandee/The-cultural-evolution-of-sustainable-societies}.

\subsection{Model structure} \label{sec:model structure}

The model includes two types of entity: agents ($i$) and patches ($j$). Table \ref{tab:Model parameters} lists the model parameters and their default values.

Agents represent cultural individuals with exosomatic (built) capital. Agents are not assigned to groups, but hold symbolic group markers denoted by colour. Markers allow agents to differentiate their strategies towards group defence, resource access and monitoring based on group membership. Group markers are not permanent and can change through mutation and imitation. Agents occupy resource patches individually. Only neighbouring agents may interact with the focal agent directly. In the local resource condition, the resource stock is contained within patches. A focal agent harvests from its local patch and its local commons, which also contains the eight neighbouring patches. 

Patches represent spatial units of renewable resource stocks (e.g. biomass). Each patch can be occupied by at most one agent, and neighbouring patches are defined using a Moore neighbourhood. The spatial environment is modelled as a square grid of $L \times L$ patches with periodic boundary conditions. The model is run in discrete time ($t$), with each step assumed to span one time unit. To explore how the level of the resource dilemma affects cooperation outcome, experiments alternate between a global and a local resource condition. In the local condition, the resource grows on each patch separately and consumers have access exclusively to patches belonging to their own commons. In the global resource condition, all consumers draw from a single commons accessible to all. All other settings remain equal between the local and global conditions.

\begin{table}[!tb]
\caption{Model parameters}
\label{tab:Model parameters}
\begin{tabularx}{\linewidth}{p{2.5cm}llX}
\textbf{Parameter} & \textbf{Symbol} & \textbf{Units} & \textbf{Value} \\ \hline
\rowcolor{gray!20} 
\textbf{Global} & & &\\\hline
Grid width/length & $L$ & patches & 41\\\hline
Initial population size & $N_0$ & agents & 500 \\\hline
\rowcolor{gray!20} 
\textbf{Agent} & $i$ & &\\ \hline
Normalization constant max harvest & $b_\text{harvest}$ & dmnl & 1 \\\hline
Half saturation constant & $h$ & kg &1 \\\hline
Normalization constant BMR & dmnl & $b_\text{bmr}$ & 1 \\\hline
Energy density & ${e}_{i}$ & J/kg & 10 \\\hline
Growth rate & $\tau$ & step$^{-1}$ & 0.5 \\\hline
Growth efficiency & $\epsilon_g$ & dmnl & 0.8 \\\hline
Assimilation efficiency & $\epsilon_a$ & dmnl & 0.8 \\\hline
Mutation rate & $\mu$ & dmnl & 0.1 \\\hline
Imitation rate & $\beta$ & dmnl & 0.1 \\\hline
Max mass & $M_\text{max}$ & kg & 100\\\hline
Birth mass fraction & $M_{\text{birth}}$ & dmnl & 0.1 \\\hline
Conflict multiplier & $\phi_{\text{fight}}$ & dmnl & 1.5 \\\hline
Migration multiplier & $\phi_{\text{move}}$ & dmnl & 1.5\\ \hline
Attrition rate & $\kappa$ & dmnl & 0.5\\ \hline
\rowcolor{gray!20} 
\textbf{Patch} & $j$ & &\\\hline
Growth rate & $r$ & step$^{-1}$ & 0.5\\\hline
Carrying capacity & $K$ & kg & 100 \\ \hline
Energy density & ${e}_{j}$ & J/kg & 10\\ \hline
Diffusion rate & $D$ & dmnl & 0.01\\ \hline
\end{tabularx}
\end{table}

\subsection{Model processes and scheduling}
The model is initialised with a random resource level between 0 and $K$, assigned to each patch separately in the local condition or to the single resource in the global condition. $N_0$ agents are spatially distributed in ten clustered groups of canonical institutional combinations randomly spread across the grid. Each cluster is assigned a random colour from a fixed palette of ten. After initialisation, the processes described below iterate once per time step. 

At the start of each time step, group boundaries are operationalised through territorial defence. A focal agent $i$ evaluates its Moore neighbourhood for possible intruders and attacks according to its group boundary trait. Agents may hold one of three strategies towards boundary defence: agents with open boundaries will not attack; agents with common boundaries will attack only neighbours with different social markers; and agents with private boundaries will attack any neighbour. Each attacker/defender dyad is resolved at most once per time step. The outcome of the conflict is determined by the mass of the two parties in conflict with a single probabilistic draw. The loser loses mass with attrition rate $\kappa$ and, if this causes death, the winning agent can move to, or reproduce on the empty patch left by the losing agent. Both defending and attacking agents incur the energetic cost of conflict due to an increased metabolic rate.  
 
The agents then harvest from either a local resource or a single global resource, depending on the experimental condition. In the local resource condition, agents may share access to their resource with neighbours depending on their resource boundary strategy. Agents with open resource access allow access to all other agents, commons agents allow access to agents who share the same group marker, and private agents exclude everyone. In addition, depending on their monitoring trait, agents can monitor the harvesting strategy of agents with whom they share access, and consequently exclude them from or include them in shared harvesting. Harvesting rates depend on both the consumer and resource variables. A consumer's maximum harvest rate ($I_{i, \text{max}}$) scales allometrically with its mass ($M_{i}$). The harvest rate ($I_i$) then follows a saturating Holling type II functional response. In the local resource condition, $E_j$ represents the total resource energy available across the agent’s local commons, defined by the cooperating set of neighbouring agents. In the global resource condition, all agents harvest from a single shared resource pool $E$. Agents may hold one of two harvesting strategies, a restrained harvester will not harvest beyond maximum sustainable yield (MSY) while an unrestrained harvester will harvest at the physical maximum rate. The harvested resource is converted into resource available for allocation with assimilation efficiency $\epsilon_a$.

Agents will migrate to empty neighbouring patches if the potential gain is greater than the cost. The potential gain is the resource level of the richest neighbouring patch minus the level on the agent's focal patch. Migration increases metabolic rate; the cost is therefore this increased rate multiplied by the single unit of time spanned by the time step.

The total assimilated resource stock available at each time step is the sum of the resource harvest in that step and the previously stored resource stock by that agent. At each time step, agents' basal metabolic rate (BMR) are calculated. These are maintenance costs that scale allometrically with body mass. Additional energetic costs are incurred when agents engage in energetically demanding activities such as conflict or migration. These activity-specific costs are modelled as multiplicative increases in metabolic expenditure relative to BMR, resulting in a field metabolic rate (FMR). The available resource is reduced by the FMR. If the assimilated resource stock is below zero after this reduction, the agent dies. 

After paying the cost of maintenance and activities, any remaining resource is allocated to somatic growth. The energy of the assimilated resource restricts the rate at which the biomass consumed can be converted to consumer body mass, with the conversion efficiency $\epsilon_g$ determining losses during synthesis. Growth is also constrained by mass: agents follow a von~Bertalanffy-type growth formulation. 

Agents are born with mass at fraction $M_\text{birth}$ of $M_\text{max}$. Those with sufficient available resources and a minimum mass may reproduce if there is space, creating offspring with identical traits, except for mutations. During reproduction, harvesting and institutional strategies and social markers may mutate separately with the same probability $\mu$. Resources passed on to offspring are removed from the parent. Any assimilated resources available after reproduction are put into storage, which follows the same von~Bertalanffy-type growth formulation as somatic growth.

Next, agents compare their mass to that of neighbouring agents. If a neighbour has a higher mass, the focal agent copies the traits of that neighbour with probability $\beta$.

Agents have an age that increases by one unit each time step. Mortality from senescence is stochastic and increases exponentially with age \cite{hansen_how_2025}. Dead consumer mass is removed from the modelled system and treated as losses to an external sink.

Resource stocks regenerate each time step following density-dependent (logistic) growth. In the local condition, each patch grows logistically independently. In the global resource condition, a single global resource energy stock follows the same logistic growth curve. $K$ is calibrated so that the system-level carrying capacity is equal in both settings. In the local resource condition, patch resources diffuse at rate $D$, such that each patch transfers a fraction $D$ of its resource per time step equally among its eight neighbours.

\subsection{Simulation experiments}
We ran two scenarios with the alternating resource dilemma structures described in section \ref{sec:model structure}. 

\subsection{Model analysis}\label{sec:model analysis}
We track all the state variables and analyse the levels of cultural selection for all behavioural strategies. George Price developed a mathematical formalism for describing genetic, non-genetic and
group selection \cite{price_selection_1970, price_extension_1972, mcelreath_mathematical_2007}. This formalism, known as the Price equation, decomposes the change in the population mean of a trait over time into selection and transmission:

\begin{equation}\label{eq:Price equation}
\Delta \bar{z} = \underbrace{\mathrm{Cov}(w, z)}_{\text{selection}} + \underbrace{\left\langle w \Delta z \right\rangle}_{\text{transmission}}
\end{equation}

where \(\bar{z}\) is the population mean trait value, \(\Delta \bar{z}\) its change over a time interval, and \(w\) is relative fitness, so that \(\bar{w} = 1\) and the usual \(1/\bar{w}\) factors drop out. Angle brackets \(\left\langle \cdot \right\rangle\) denote an average over the units indicated, weighted by relative group size where the average is taken over groups. The selection term is positive when trait values and fitness covary positively, driving \(\bar{z}\) upward, and is known as the selection differential (\(S\)). The transmission (\(T\)) term reflects systematic differences between the trait value of ancestors and their descendants, such as through mutation. \textcite{kline_designing_2018} propose analysing group-level cultural adaptations for resource use by decomposing the selection term of equation \eqref{eq:Price equation} into group and individual selection. We use the following formulation:

\begin{equation}\label{eq:Selection differential}
S =
\underbrace{\mathrm{Cov}(w_g, z_g)}_{\text{group selection}} +
\underbrace{\left\langle \mathrm{Cov}(w_{ig}, z_{ig}) \right\rangle}_{\text{individual selection}}
\end{equation}

where \(z_g\) is the group mean trait value, \(w_g\) is the group mean relative fitness, \(w_{ig}\) is the individual relative fitness and \(z_{ig}\) is the individual trait value. Individual fitness is scaled by the population mean rather than by the group mean, which is what makes the decomposition exact. Group means and among-group covariances are likewise weighted by group size, so that the group and individual components sum exactly to the population-level selection term.

We analyse the levels of selection using equation~\eqref{eq:Selection differential} and the MLS1 method presented by \textcite{hermsen_emergent_2022}. The change in the population mean of a trait is tracked over time interval $\Delta t = t_2 - t_1$. Each agent is assigned a unique ID at $t_1$, becoming an ancestor, and at $t_2$ the number of descendants per ancestor is counted. Descendants are agents who inherit traits of an ancestor, whether biologically (offspring) or culturally (imitation). An agent alive at both $t_1$ and $t_2$ is its own ancestor and counts as one of its own descendants, so that absolute fitness combines survival, reproduction and imitation. Absolute fitness ($W_i$) is thus defined as the number of descendants of agent $i$ between time intervals, and relative fitness as $w_i = W_i / \bar{W}$, with $\bar{W}$ the population mean absolute fitness at that interval; an agent that dies without reproducing and without being imitated therefore has $W_i = 0$. We compute the components of the cultural selection differential over each time interval for behaviours towards harvesting, group boundary, resource boundary and monitoring.

The model includes mutation of behavioural strategies, contributing to the transmission term, which we do not analyse here as our interest is in the relative strength of selection at the two levels. The two components of \(S\) therefore account only for the selective part of \(\Delta \bar{z}\). In addition, note that there are numerous conflicting definitions of (cultural) fitness \cite{ramsey_is_2017}, and other definitions are possible. For example, cultural fitness might be considered relevant to wealth or mass.

\section{Results}
\begin{figure*}[t]
    \centering
    \includegraphics{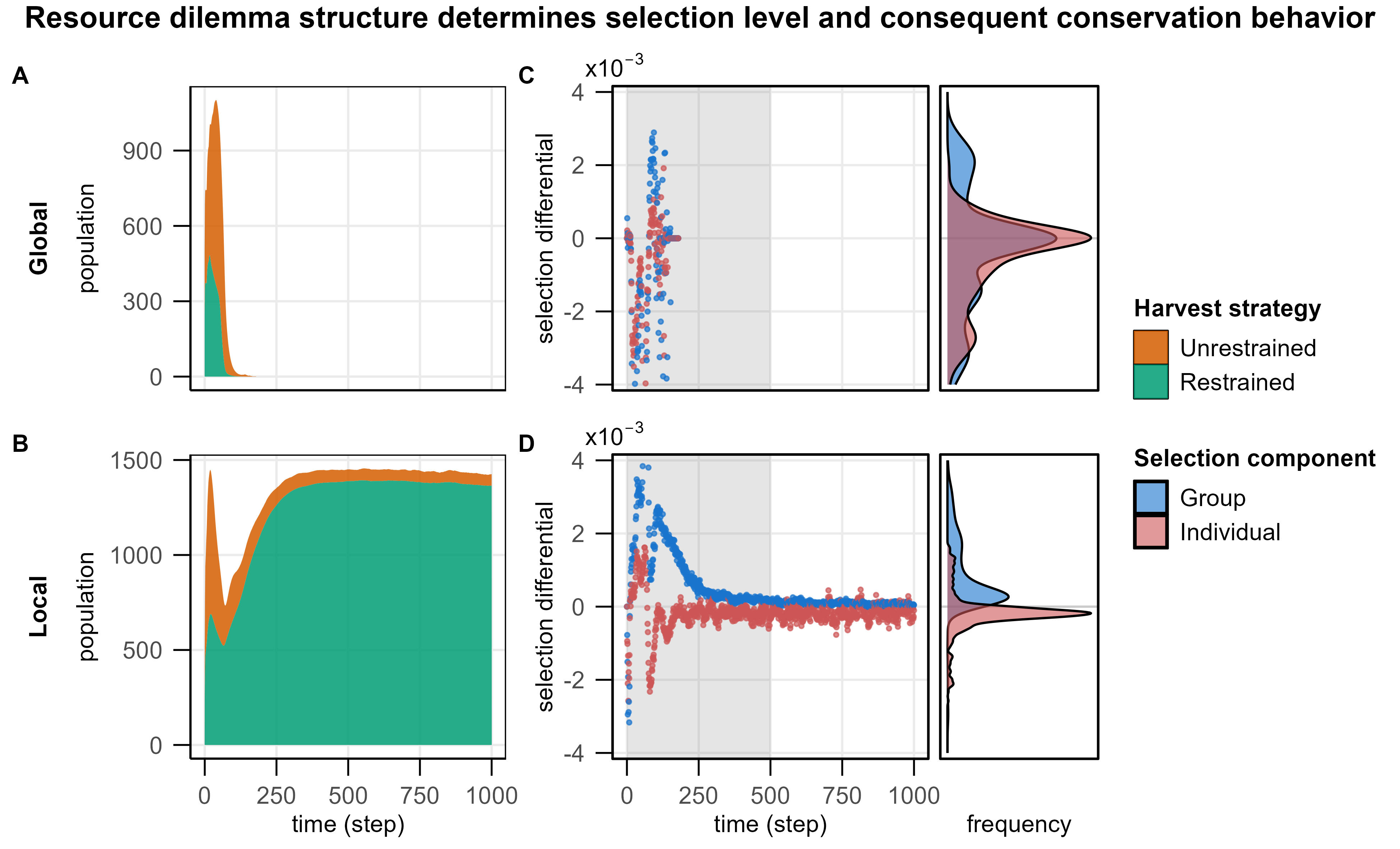}
    \caption{(A, B) Means for number of agents with restrained (green) and unrestrained (orange) harvesting strategies over time and (C, D) the average strength of selection for harvesting restraint at the group (blue) and individual (red) level over time in global and local resource conditions. The rotated histograms on the right-hand side show distributions of the selection measurements based on the first 100 steps of the global resource condition and the first 500 steps of the local resource condition, indicated with the gray background. These parts of the simulation held the most significant portion of change to the population, after which it collapsed or stabilized.}
    \label{fig:Local and global harvest strategies and their selection decompositions}
\end{figure*}

\begin{figure*}[t]
    \centering
    \includegraphics{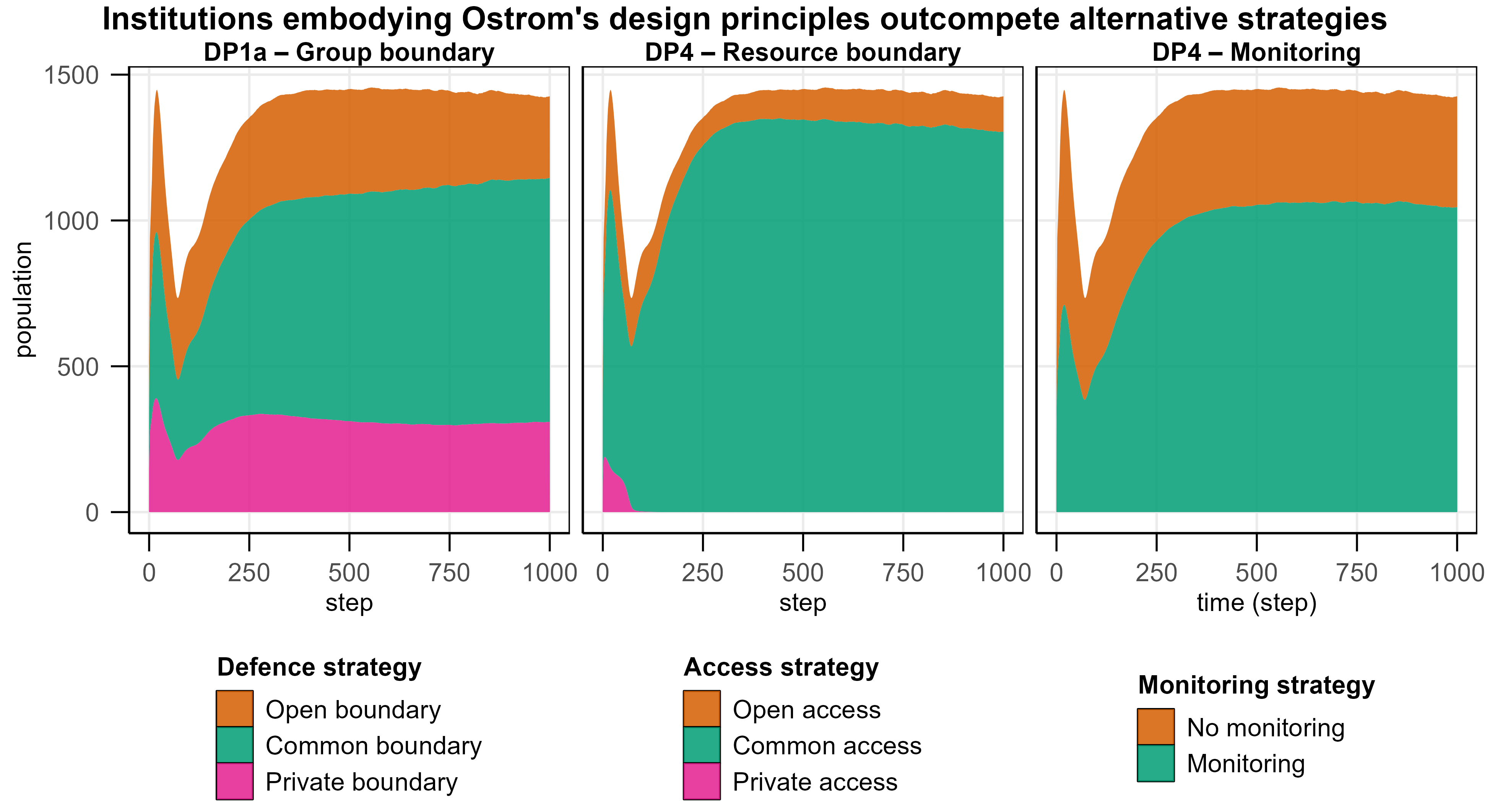}
    \caption{Stacked mean frequency of each institutional strategy in the population over time.}
    \label{fig:Local institutional strategies}
\end{figure*}

\begin{figure*}[t]
    \centering
    \includegraphics{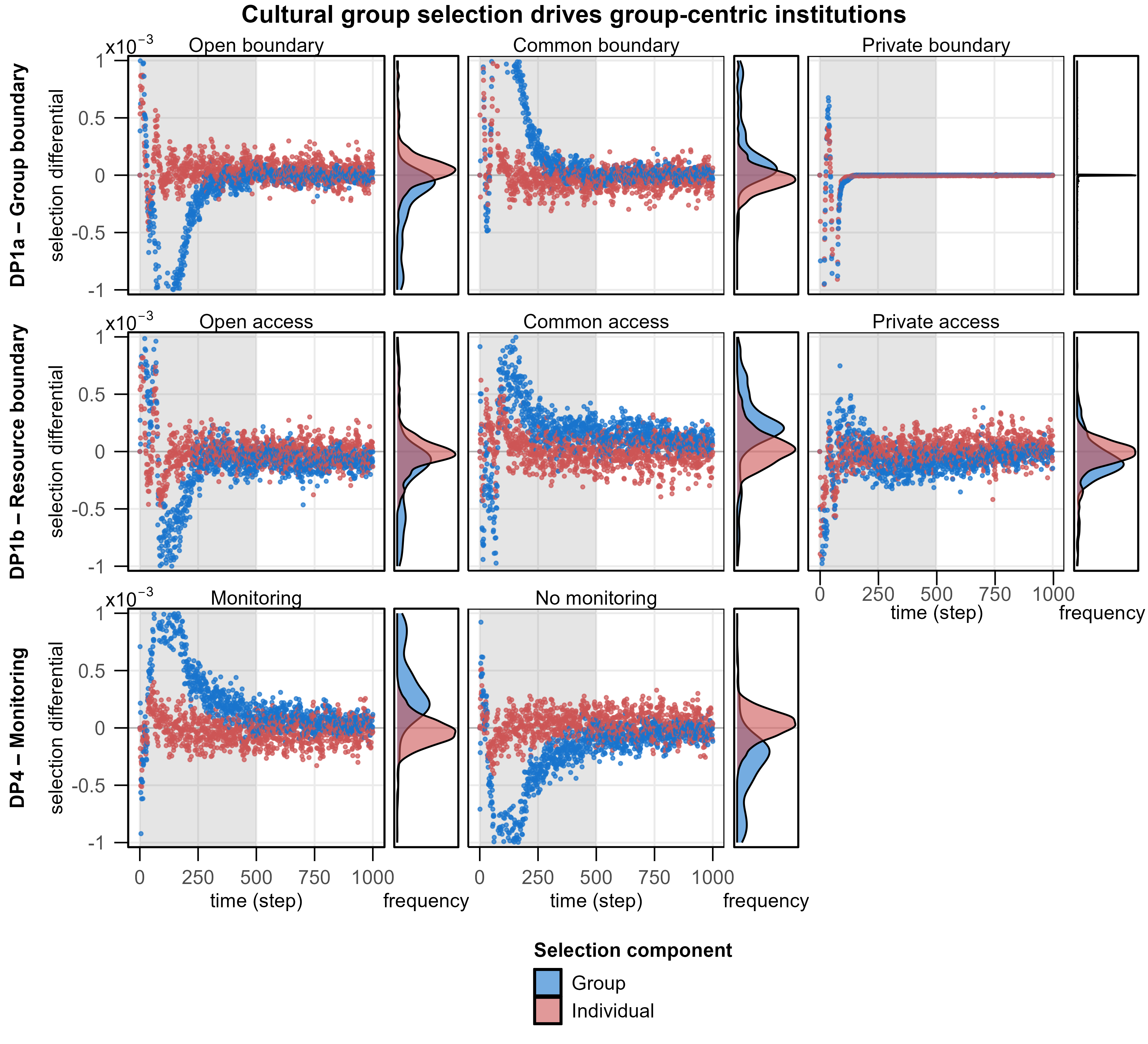}
    \caption{Selection for each institutional strategy over time, decomposed into individual and group levels for group, resource and monitoring strategies by type. The rotated histograms on the right-hand side show distributions of the measurements based on the first half of the simulation, indicated with the gray background. This part of the simulation held the most significant portion of change to the population, after which it stabilized. Components of selection among strategies are zero-sum within each row, so gains for one category are exactly offset by losses across the others, as agents hold only one strategy per institutional type.}
    \label{fig:Selection decomposition of local institutional strategies}
\end{figure*}

Simulation results follow consistent patterns under the different simulation conditions. In all experimental conditions, agents grow in both size and number until the grid space is filled and the resource can no longer support growth. The model results align with those of \textcite{waring_coevolution_2017} in that resource collapse leads to either population collapse or selection of those with harvesting restraint. However, our results reveal stark differences in social evolution based on the nature of the resource. In the local condition, populations reach $t=1000$ in 98\% of the simulation runs. In these runs, total population mass peaks on average at $t=25$ ($\pm$ 6.9 SD) of which a mean 40\% ($\pm$ 12\% SD) then reaches $t=100$. This fraction of the population restrains its harvest and manages to defend its property from over-harvesting, preventing local resource collapse. Those that do not, however, collapse their immediate environment and subsequently die out. After such local resource collapses, the restrained harvesters who survived expand slowly when resource diffuses into uninhabited and depleted territory. In the global resource condition, resource is depleted globally, leading to a total population collapse in all simulations at a mean time $t=123$ ($\pm$ 9.9 SD). The following sections first compare simulation outcomes of the two resource dilemma settings and then explore the evolution of institutional strategies in the local resource dilemma.

\subsection{Resource dilemma structure and levels of selection}
Selection for harvesting restraint is determined by the structure of the resource dilemma, as seen in figure \ref{fig:Local and global harvest strategies and their selection decompositions}. When the resource is globally accessible (so that the resource dilemma is above the level of selection), resource and consumer populations collapse following selection for unrestrained harvesting, as expected under the CMLS framework. Meanwhile, the local resource dilemma condition selects for harvesting restraint, and consequently consumers persist. 

In the global resource dilemma condition, the decomposition reveals that both individual and group selection for restrained harvesting strategies are predominantly negative (panel C of figure \ref{fig:Local and global harvest strategies and their selection decompositions}). Over the entire simulation run, the group component of the cultural selection differential was negative during 71\% of time intervals while the individual component was negative during 78\% of intervals, indicating that both group and individual selection against harvesting restraint drive resource and consumer collapse in a global resource dilemma.

A different pattern emerges when the resource dilemma is structured locally, shown in panels B and D of figure \ref{fig:Local and global harvest strategies and their selection decompositions}. The decomposition of the cultural selection differential shows that its group and individual components for restrained harvesting strategies hold largely opposing signs. Over the first half of the simulation run, the group component of selection was positive during 96\% of the time intervals. In that same period, the individual component was positive during 12\% of time intervals. However, the total selection differential remains positive for harvesting restraint. Both group and individual selection converge to zero at around 300 time steps when restrained harvesting has become the single most dominant strategy. Thus, while individual selection for restrained harvesting is negative, this is overcome by a stronger positive group selection for harvesting restraint.

\subsection{The evolution of institutions}
Since both the resource and consumers collapse in the global resource experimental condition, this section covers the results of experiments in the local resource dilemma condition, further exploring the evolution of local institutions in the simulations. 

Figure \ref{fig:Local institutional strategies} displays the evolution of each institutional strategy over time in the local resource dilemma condition. The panels show that for each institution type, those embodying Ostrom's design principles become the dominant strategy in the population. Meanwhile, small fractions of the population with alternative strategies co-exist or are reintroduced through random mutation, while only the private boundary defence strategy is completely suppressed.

Selection for each institutional strategy is again decomposed into individual and group selection and displayed in figure \ref{fig:Selection decomposition of local institutional strategies}. For each strategy, the cultural selection differential and its group and individual components converge to zero at around half of the simulation period when one combination of strategies is held by most of the population without changing. Group selection drives all group-centric institutions, while individual selection remains relatively weak. During the first half of the simulation run, group selection for common group boundaries is positive during 82\% of the time intervals, group selection for common access is positive during 87\% of the time intervals and group selection for monitoring is positive during 94\% of time intervals. Note that individual and group benefits of institutional strategies depend on the combination of institutional settings rather than on each setting separately. For instance, without monitoring and consequent enforcement, common property creates a within-group resource dilemma where individuals benefit from unrestrained harvesting but the group benefits from collective restraint. These results support the hypothesis that institutions following Ostrom's design principles for sustainable common-pool resource management emerge through a mechanism of cultural multilevel selection.

\section{Discussion}
Presented here is a stock-flow-consistent agent-based model of generic consumers competing over renewable resources under physical constraints. Agents share the same physical properties and constraints, but variation, selection, and replication of harvesting strategies, institutional strategies, and social markers allow the cultural evolution of competing property rights and group formation. Between experimental conditions, resource properties remain equal, but access structure is altered. Results show that when all agents have access to a single resource, agents with maximal harvesting strategies out-compete agents with restraint in the short term, but reliably collapse resource and consumer populations in the long term. When the scale of resource access is at the level of the agents, allowing them to individually form property rights, agents with restraint towards harvesting will prevent local resource collapse and out-compete unrestrained harvesting strategies. Decomposition of the levels of selection shows that when the level of the resource dilemma is the same as the level of selection, both group and individual selection will select against harvesting restraint. When the resource dilemma is structured below the level of selection, group selection for group-functional traits overcomes individual selection for individual traits, allowing those groups to survive. 

Biophysical processes and environmental conditions explain the emergence of different cooperative institutional and harvest strategies. A single open-access renewable resource drives agents towards individual (short-term) maximization of consumption. Note that in this model, agents have no cognition, and collective behaviours may only emerge endogenously, so that given a global pool of resources, feedback on resource status only becomes significant through collective harvesting when a tragedy finally occurs. In contrast, local resources provide a more direct biophysical feedback, allowing agents with sustainable harvesting strategies to proliferate under some parameter settings. 

The cultural evolution of strategies is shaped by the biophysical properties of agents and the interactions between agents and their environment. The evolution of institutional strategies, facilitated by social markers, allows a similar dynamic to play out at the group level. A classical tragedy of open access plays out when agents are granted access to neighbours' local resources but do not reciprocate and thus gain higher fitness by over-harvesting \cite{anderies_sustaining_2016, hardin_tragedy_1968}. Meanwhile groups of agents that keep local resources in common can proliferate. Contrary to \textcite{waring_coevolution_2017}, common property is the most dominant strategy in the long term when initial assortment is high. Our model recapitulates Ostrom's findings when clusters of agents hold common property access norms, creating conditions that favor harvesting restraint, and maximizing local resource use in the long term. 

Evolution suppresses conflict in two ways. First, those involved in conflict are more prone to perish directly. Second, metabolic throughput spent on conflict cannot be spent on growth or reproduction. Agents with a private group boundary strategy attack everyone, and soon go extinct. With high initial assortment, however, group-centric boundary defence is the long-term dominant strategy. \textcite{zefferman_evolutionary_2015} provide a theory for how large-scale human warfare can be explained through group-structured cultural selection. \textcite{turchin_warfare_2010} hypothesize that this mechanism of warfare caused the rise of large complex societies by increasing cooperation. Although its warfare dynamics are highly stylised, our model further demonstrates the emergence of group structures through the cultural selection of groups of agents with shared markers and a common group boundary strategy.

This work highlights the notion that societies and ecosystems cannot be understood independently because they interact and co-evolve over time \cite{haberl_contributions_2019}. Societies both shape and are shaped by their biophysical environment; sustainability science thus requires integrating the study of socioeconomic processes with biophysical processes. Our model does so by integrating concepts from sociometabolic research with commons and CMLS theory. It also refines the multilevel evolutionary framework for sustainability analysis \cite{waring_multilevel_2015}. Contrary to what this model suggests, global resource dilemmas need not inevitably lead to system collapse. Instead, our results add to the evidence that intentional cultural adaptation at the group level is needed to overcome such dilemmas \cite{richerson_agentic_2023, wilson_multilevel_2023, waring_characteristic_2023}. Fortunately, evidence for successful global environmental protection exists, with the preservation of the ozone layer as a prominent example \cite{chipperfield_detecting_2017}.

Our implementation of sociometabolic theory is also stylized and leaves many avenues for future improvement. On the consumer side, human societies might not follow the growth trajectories represented here. Moreover, we presented two competing harvesting strategies. In reality, individual agents' harvesting strategies are not fixed but rather evolve over time. \textcite{burger_extra-metabolic_2017}, for instance, show variation in densities and individual energy use in human societies. This could be implemented in our model by encoding harvesting strategy as an evolvable trait. Finally, fitness is defined here as the number of offspring generated by an agent. Burger et al. argue that fitness should be defined as the energetic equivalent of surviving offspring per generation \cite{burger_universal_2021}. This equal fitness paradigm might be useful for linking evolutionary processes across sociometabolism and culture.

On the resource side, we modelled flows of one renewable resource that fuels competing societies. In reality, societies depend on flows and stocks of many renewable and non-renewable resources, uses of which are constrained not only by source-side constraints like resource depletion but also sink-side constraints like pollution. Indeed, the real-world metabolic rift with nature is fuelled predominantly by burning finite fossil fuel stocks created on timescales far longer than those of human societies \cite{burger_extra-metabolic_2017, schramski_human_2015}, and by the resulting pollution. Future research might explore how the dynamics revealed in our results shift when societies can also access non-renewable resources and face sink-side constraints. A range of resources could be included (e.g., renewable flows such as wind, hydro or solar, or finite stocks such as fossil fuels), as well as material stocks (e.g., building materials or critical minerals), sinks (e.g., the carbon sink of the atmosphere), and pollution (i.e. the reduction of biological capacity due to resource contamination). 

Calibrating the expanded model with real-world data would be another step forward. Most sociometabolic data, especially for flows, are currently available only on the national level. Data for material stocks (e.g. materials in artifacts such as buildings, infrastructures and machinery at a specific point in time) are increasingly becoming available \cite{haberl_weighing_2025}, but they have not yet been linked to comprehensive flow accounts such as those in material flow analysis. Scholars have argued for the creation of, and use cases for, such datasets \cite{burger_extra-metabolic_2017, haberl_social_2016}, and linking them to complexity science and political ecology \cite{haberl_resilience_2026}. We propose that analysing comprehensive stock and flow data in relation to social and institutional dynamics may enable the study of evolutionary pathways towards sustainable sociometabolism.

\section{Conclusions}
Human groups that collectively maintain their local environment on land held in common have been abundant and persistent in cultures around the world \cite{cox_review_2010, mbembe_communaute_2023, kimmerer_braiding_2013, baragwanath_collective_2020}. In this work we integrated cultural multilevel selection theory with commons scholarship and sociometabolic research in a stylized agent-based model that provides mechanistic explanations for the emergence and persistence of such societies. Consistent with the CMLS framework for sustainability, our model suggests that resource dilemma structure is the main determinant of the evolutionary dynamics. When the dilemma is set below the level of selection, the presence of institutions that manage common property and resource boundaries enables the emergence of cultural groups that sustainably manage shared resources. Moreover, our results show that, in an energetically consistent model of consumers and a renewable biomass resource, cooperating agents with harvesting restraint may co-exist with, or out-compete, non-cooperating agents with unrestrained harvesting by collectively protecting and managing their common resource pool. 

\subsection*{Acknowledgments}
We are grateful to Robert Harmsen for co-supervising TZ's Master thesis project, from which this study emerged. 

\subsection*{Data Availability}
The Netlogo model and R analysis and plot scripts needed to reproduce all figures and results are available through \url{https://anonymous.4open.science/r/Cultural_multilevel_selection_of_institutions_for_sustainable_social_metabolism-0D42/}.

\printbibliography

\onecolumn

\appendix
\section{ODD} \label{Appendix A}

\section{Additional simulation treatments} \label{Appendix C}
The following section provides a complete model description following the ODD (Overview, Design concepts, Details) protocol for describing individual- and agent-based models (Grimm et al., 2006), as updated by Grimm et al. (2020). The ODD is designed to be hierarchical, starting with the most important features, followed by design rationale and details. 

\subsection{Purpose and patterns}
The purpose of this model is to explain how sustainable resource use emerges in a stock-flow-consistent community of generic cultural consumers competing over renewable resources. The mechanism driving the explanation is the cultural multilevel selection for Ostrom’s institutional design principles. These principles have been empirically found to contribute to group-level cultural adaptations and in many successfully managed shared-resource systems. The model represents a bio-physically realistic abstraction of human societies with varying institutional and consumption strategies, harvesting a renewable resource to see how restraint in resource use may outcompete unsustainable consumption and consequent growth. The model reveals if any patterns of group level conservation behaviour and consequent survival patterns are the consequence of cultural multilevel selection. Groups and the institutional rules that maintain them may emerge and dissolve as a consequence of their fitness. The structure of the resource dilemma is altered between experiment treatments. Results show that when all agents have access to a single resource, agents with maximal harvesting strategies out-compete agents with restraint in the short term, but reliably collapse resource and consumer populations in the long term. When the level of resource access is at level of the agents, allowing them to collectively form property rights, groups of agents agents with restraint towards harvesting will prevent local resource collapse and out-compete maximum harvesting strategies. The ultimate purpose of this model is to estimate evolutionary trajectories of resource management systems. 

\subsection{Entities, state variables and scales}
\begin{figure}[htbp]
    \centering
    \includegraphics[width=100mm]{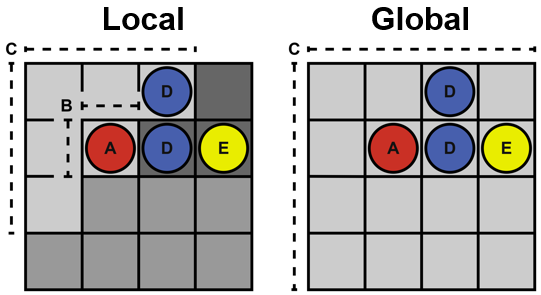}
    \caption{Model spatial structure and resource settings. Agents (circles) occupy resource patches (squares) individually. Only neighbouring agents (D) may interact with the focal agent (A) directly. Agent colors denote social makers. In the local resource setting (left), resource stock (shades of grey) is contained within patches. A focal agent harvests from its local patch (B), and its local commons (C), which also contains the eight neighboring patches. In the global resource setting (right), resources are a single stock (C) accessible to all. All other settings remains equal between the local and global setting.}
    \label{fig:ODD:Model spatial structure}
\end{figure}

The model includes two types of entity: agents ($i$) and patches ($j$). The state variables characterizing these entities are listed in Table \ref{tab:ODD:Model variables}. \ref{tab:ODD:Model parameters} lists the model parameters and their default setting, with energy in joules and mass in kilograms. Figure \ref{fig:ODD:Model spatial structure} displays the model spatial structure.

Agents represent cultural individuals with exosomatic (built) capital. Agents are not assigned to groups, but hold symbolic group markers denoted by colour. Markers allow agents to differentiate their strategies towards group boundaries, resource boundaries and monitoring based on group membership. Group markers are not permanent and can change through mutation and imitation. All behaviours are individual or between neighbouring agents, so any collective patterns are emergent. 

Patches represent spatial units of renewable resource stocks (e.g. biomass). Each patch can be occupied by at most one agent, and neighbouring patches are defined using a Moore neighbourhood. The spatial environment is modelled as a square grid of $L \times L$ patches with periodic boundary conditions. The model is run in discrete-time, with each step ($t$) representing one unit of time. To explore the effect of the level of the resource dilemma on the cooperation outcome, experiment treatments alternate between a global and a local resource treatment. In the local treatment, resources grow on each patch separately, and consumers have access exclusively to patches belonging to their own commons. In the global resource treatment, all consumers draw from a single commons accessible to all. All other settings remains equal between the local and global setting.

\begin{table}[htbp]
\caption{Model parameters}
\label{tab:ODD:Model parameters}
\begin{tabularx}{\linewidth}{p{4cm}p{1cm}p{1cm}p{7.5cm}X}
\textbf{Parameter} & \textbf{Symbol} & \textbf{Units} & \textbf{Definition} & \textbf{Value} \\ \hline
\rowcolor{gray!20} 
\textbf{Global} & & & &\\\hline
Grid width/length & $L$ & patches & Width and length of the square grid & 41 \\\hline
Initial population size & $N_0$ & agents & Number of agents at the start of the simulation & 500 \\\hline
\rowcolor{gray!20} 
\textbf{Agent} & $i$ & & & \\ \hline
Normalization constant max harvest & $b_\text{harvest}$ & - & Scaling parameter harvest & 1 \\\hline
Half saturation constant & $h$ & kg & The resource availability at which half of the
maximum harvest rate is reached & 1 \\\hline
Normalization constant BMR & $b_\text{bmr}$ & - & Scaling parameter BMR &  1 \\\hline
Energy density & ${e}_{i}$ & J\,/kg & Amount of energy per unit of consumer mas & 10 \\\hline
Growth rate & $\tau$ & step$^{-1}$ & The rate at which an agent approaches its
asymptotic maximum mass & 0.5 \\\hline
Growth efficiency & $\epsilon_g$ & - & The efficiency at which consumer energy is
converted into body mass & 0.8 \\\hline
Assimilation efficiency & $\epsilon_a$ & - & The efficiency at which harvested resources are
converted into consumer resources & 0.8 \\\hline
Mutation rate & $\mu$ & - & Chance offspring mutate their behavioural
strategies and social marker & 0.1 \\\hline
Imitation rate & $\beta$ & - & Chance agents copy the behavioural strategy of
neighbouring agents with a higher mass & 0.1 \\\hline
Max mass & $M_\text{max}$ & kg & Maximum size of an agent in mass & 100 \\\hline
Birth mass fraction & $M_{\text{birth}}$ & - & Initial mass of offspring as a fraction of its parent & 0.1 \\\hline
Conflict multiplier & $\phi_{\text{fight}}$ & - & Factor scaling metabolic rate during conflict & 1.5 \\\hline
Migration multiplier & $\phi_{\text{move}}$ & - & Factor scaling metabolic rate during migration & 1.5 \\ \hline
Attrition rate & $\kappa$ & - & - & 0.5 \\ \hline
\rowcolor{gray!20} 
\textbf{Patch} & $j$ & & & \\\hline
Growth rate & $r$ & step$^{-1}$ & Intrinsic rate of biomass increase & 0.5 \\\hline
Carrying capacity & $K$ & kg & Maximum size of resources in mass & 100 \\ \hline
Energy density & ${e}_{j}$ & J\,/kg & Amount of energy per unit of resource mass & 10 \\ \hline
Diffusion rate & $D$ & - & Fraction of patch resources redistributed to
neighboring patches per time step & 0.01\\ \hline
\end{tabularx}
\end{table}

\begin{table}[htbp]
\caption{Model state variables}
\label{tab:ODD:Model variables}
\begin{tabularx}{\linewidth}{p{2cm}p{1.5cm}p{1cm}p{8cm}X}
\textbf{Variable} & \textbf{Symbol} & \textbf{Units} & \textbf{Definition} & \textbf{Range} \\\hline
\rowcolor{gray!20} 
\textbf{Agent} & $i$ & & &  \\ \hline
Harvesting strategy & - & - & Binary trait determining if agents will harvest beyond sustainable yield & (Restrained, Unrestrained) \\\hline
Group boundary & - & - & Ternary trait determining which neighbours an agent will initiate conflict against & (Open, Common, Private) \\\hline
Resource boundary & - & - & Ternary trait determining which neighbours an agent will share access to resources with & (Open, Common, Private) \\\hline
Monitoring strategy & - & - & Binary trait determining if an agent will monitor shared harvesting preferences and sanction (exclusion from shared harvesting) & (Non-monitor, Monitor) \\\hline
Energy & $E_i$ & J & Current amount of consumer energy & (0 - $\infty$) \\ \hline
Stored mass & $M_i^{\text{stored}}$ & kg & Current amount of mass available for use & (0 - $\infty$) \\\hline
Structural mass & $M_i^{\text{structural}}$ & kg & Current consumer mass & ($M_{\text{birth},i}$ - $M_\text{max}$) \\\hline
Maximum harvest rate & $I_{\text{max}, i}$ & J\,/step & Upper limit on harvest rate, scaling allometrically with body mass & (0 - $\infty$) \\\hline
Harvest rate & $I_{i}$ & J\,/step & Realized amount of resource energy harvested
from a resource by an agent in a given time step, limited by maximum harvest rate and
resource availability & (0 - $I_{\text{max}, i}$) \\\hline
Conflict? & - & - & Current state of territory defence & (true, false) \\\hline
Migration? & - & - & Current state of migration & (true, false) \\\hline
Basal metabolic rate & $BMR_{i}$ & J\,/step & Amount of energy expended per time step to maintenance, scaling allometrically with body mass & (0 - $\infty$) \\\hline
Field Metabolic rate & $FMR_{i}$ & J\,/step & Amount of energy expended per time step to
both maintenance and activities & (0 - $\infty$) \\\hline
Age & $A_{i}$ & steps & Number of simulation time steps the agent has
survived & (0 - 100] \\\hline
\rowcolor{gray!20} 
\textbf{Patch} & $j$ & & & \\\hline
Energy & $E_j$ & J & Current resource energy & (0 - $K \cdot {e}_{j}$) \\ \hline
Mass & $M_j$ & kg & Current resource mass & (0 - $K$) \\ \hline
\end{tabularx}
\end{table}

\subsection{Process overview and scheduling}
The model is set in discrete timesteps. After initialization, the model processes the following submodels once each timestep (see section \ref{sec:ODD:submodels} for the details of each submodel):

\begin{enumerate}
    \item Agents execute their \textbf{territory defense} submodel in which they pair up with opponent agents and engage in conflict, the outcome of which affects the losing agents' \textit{mass}. The agent variable \textit{conflict?} is updated during this timestep, which is used during the allocation submodel to calculate activity cost. The order of the agents executing this submodel is random each timestep and the submodel is executed for each possible pair combination once per timestep.
    \item Agents execute their \textbf{harvest} submodel where they harvest resources (mass and energy) from the patches to which they have access. Agent variables \textit{maximum harvest rate, harvest rate, ingested mass} and patch variables \textit{mass} are updated during this submodel. The order of the agents executing this submodel is random each timestep.
    \item Agents execute their \textbf{migration} submodel where they migrate if an empty neighbouring patch is available and the energetic benefits of migration outweigh its costs. The agent variable \textit{migration?} is updated during this timestep, which is used during the allocation submodel to calculate activity cost. The order of the agents executing this submodel is random each timestep.
    \item Agents execute their \textbf{allocation} submodel, where harvested resources are allocated to processes in the following sequential order:
    \begin{enumerate}
        \item Agent \textbf{BMR} (basal metabolic rate) is updated based on its body mass.
        \item The BMR is then increased by \textbf{activity cost} from conflict and migration, yielding a field metabolic rate (FMR). Available resources are reduced by resources spent on FMR.
        \item Resources are then spent on \textbf{growth}, which is constrained by physiology (limits to the rate at which new body mass can be synthesized).
        \item If agents size and and remaining resource are above reproduction thresholds and there are empty neighboring patches available, agents will continue to\textbf{reproduction} of an offspring agent into the empty neighboring patch which inherits the traits of the parent.  
        \item Offspring agents' traits have a chance of \textbf{mutation}, where they may randomly change during reproduction.
        \item Any assimilated resources available after reproduction are put into \textbf{storage}, which follows the same von~Bertalanffy-type growth formulation as somatic growth.
    \end{enumerate}
    The order of the agents executing this submodel is random each timestep.
    \item Agents execute their \textbf{imitation} submodel where they copy behavioral strategies of a neighboring agent with a greater mass at an imitation rate. Agent variables \textit{harvesting stratety, group boundary, resource boundary and monitoring strategy} may be affected during this submodel. The order of the agents executing this submodel is random each timestep.
    \item Agents execute their \textbf{age} submodel where agent variable \textit{age} is incremented and agents die with increasing probability as age increases. The order of the agents executing this submodel is random each timestep.
    \item Patches execute the \textbf{resource regrowth} submodel where resource stocks regenerate, affecting patch variable \textit{mass}. The order of the patches executing this submodel is random each timestep.
    \item Patches execute the \textbf{resource diffusion} submodel where resources diffuse into neighboring patches, affecting patch variable \textit{mass}. The order of the patches executing this submodel is random each timestep. This submodel is disabled in the global resource setting.
\end{enumerate}

\subsection{Design concepts}
\subsubsection{Basic principles}
This model incorporates insights from metabolic ecology into the CMLS framework of \textcite{waring_coevolution_2017}, following the recommendations of \textcite{sibly_representing_2013}, to investigate how institutions regulating resource use as described by \textcite{ostrom_governing_2015} evolve under explicit biophysical constraints.

Like \textcite{waring_coevolution_2017}, the model employs endogenous group structures, meaning that groups are not represented formally in the model in any way, and can only be measured. Instead, individuals hold traits including group marker traits through which they differentiate interactions with neighboring agents. Social markers themselves evolve as agent may copy them them strategically from neighbors through payoff biased learning. In this way our model has endogenous group formation and interactions (see \ref{fig:ODD:Selection method}). This method permits the analysis of the strength of selection at both individual and group levels presented in section \ref{sec:model analysis}.

\subsubsection{Emergence}
\begin{figure}[H]
    \centering
    \includegraphics[width=0.5\linewidth]{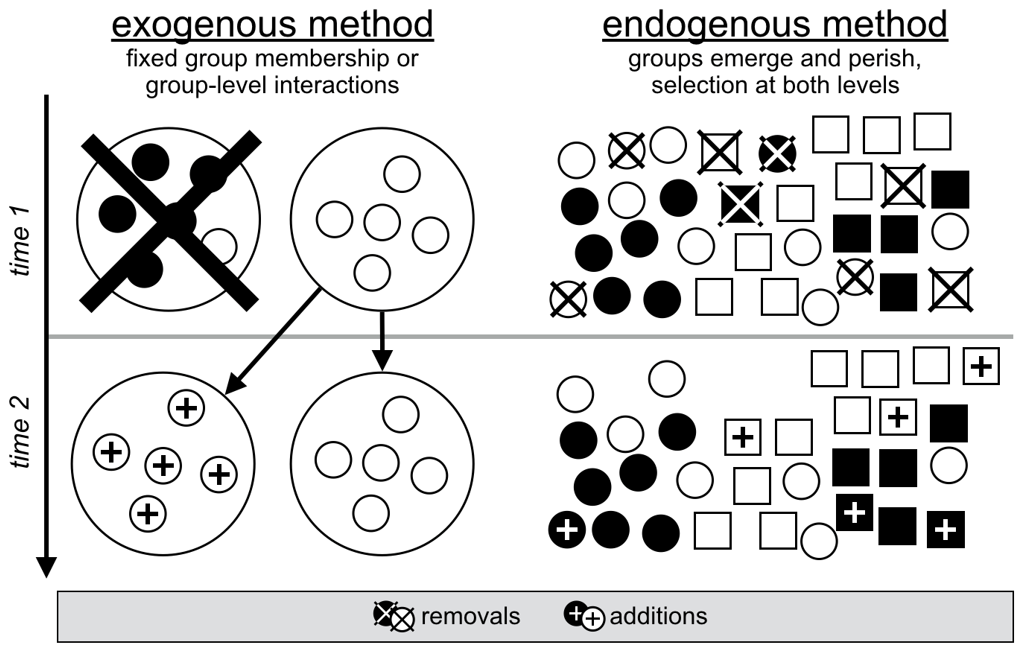}
    \caption{A graphical comparison of the endogenous and exogenous methods of modeling group selection (from \textcite{waring_coevolution_2017})}
    \label{fig:ODD:Selection method}
\end{figure}

The key outcomes of the model are system level (un)sustainability, group patterns and the evolution of institutional strategies; these results emerge from which individual strategies of agents towards each other and their environment are selected for. System level sustainability emerges from the combined effects of harvesting strategies of individual agents. Since the model employs endogenous multilevel selection mechanisms (see \ref{fig:ODD:Selection method}), groups may emerge and perish as a consequence of their fitness. 

\subsubsection{Adaptation}
The model agents have two adaptive behaviors: migration and imitation. During migration, agents can move to empty neighboring patches if the potential gain is greater than the cost. The potential gain is the amount of resources on the neighboring patch with the most resources, minus the amount of resources on the agents’ focal patch. Migration causes an increased metabolic rate, the cost is therefore the increased metabolic rate due to migration. During imitation, agents compare their total mass ($M_i$), to that of neighboring agents. If a neighbor has a higher mass, the focal agent copies the traits of that neighbor with probability $\beta$.

\subsubsection{Objectives}
The model includes two behaviors that uses direct objective-seeking: migration and imitation. The objective measure used by agents is mass, as agents use resources for body maintenance, growth and reproduction, and resource and body size are measured in mass. For migration, this means agents will move to patches if this allows a potential increase in resource mass uptake. For imitation, this means agents may copy agents with a higher body mass.

\subsubsection{Learning}
Agents do not have memory and do not change their behavior as a consequence of their experience. The only learning is payoff-biased social learning described in section \ref{par:ODD:Imitation}, which represents cultural inheritance within and between generations. 

\subsubsection{Prediction}
Prediction is not implemented.

\subsubsection{Sensing}
The model does not incorporate mechanisms of sensing and agents are assumed to have perfect information of the state variables of entities with whom they interact.

\subsubsection{Interaction}
Agents interact with other agents and resources. In the local resource treatment, all interactions are direct. Agents can interact with the patch they stand on, neighboring patches and neighboring agents (neighbors are defined through a Moore neighborhood). 

In the global resource treatment, interactions with resources are mediated. 

\subsubsection{Stochasticity}
The model employs stochasticity in multiple processes for two reasons.

Stochasticity is first used to create variability in initial conditions so that results are not artifacts of a single initial configuration. Resources (in both the local and global resource treatments) are initialized with a mass drawn from a uniform distribution between 0 and $K$. Agents are placed at random locations on the grid in the unstructured experiment treatment and clusters of agents are placed at random locations on the grid in the structured experiment treatment.

Stochasticity is then used to simplify sub-models. Conflict outcomes are are resolved probabilistically rather than deterministically, with agent body mass used as a proxy for combat ability and the probability of winning being the ratio of the two contenders' mass (see equation \ref{eq:ODD:Conflict draw}. This represents the determinants of conflict outcomes not explicitly modeled (e.g. terrain, tactics, technology) which introduce variance around the outcome expected from mass alone. During agent reproduction, each trait may mutate independently with probability $\mu$. This represents imperfect trait transmission between generations (e.g. genetic mutation in biological systems or innovation in cultural systems). Agent imitation is also probabilistic, where agents choose the neighboring agent (if any) with the highest body mass (and higher than the focal agents' mass) and copy their traits independently with probability $\beta$, representing payoff-biased social learning with imperfect trait transmission within generations (e.g. imperfect social information, inconsistent adoption, alternative payoff metrics). Agent body mass is again used here as a proxy for perceived success. Finally, agent death due to senescence increases logistically with probability $P_{death}$ (see equation \ref{eq:ODD:Age}), representing unmodeled sources of variation around death expected from age alone (e.g. accidents or health).

\subsubsection{Collectives}
All behaviors are individual or between neighboring agents, so any collective patterns are emergent. Model analysis (see section \ref{sec:ODD:observation}) of cultural multilevel selection is performed by assigning agents to groups based on their social marker (their color).

\subsubsection{Observation}\label{sec:ODD:observation}
We tracked all the state variables of the model. In addition, we analyze the levels of cultural selection for all behavioral strategies. George R. Price developed a mathematical formalism for describing genetic, non-genetic and group selection \cite{price_selection_1970, price_extension_1972, mcelreath_mathematical_2007}. This formalism, known as the Price equation, decomposes the change in the population mean of a trait over time into
selection and transmission:

\begin{align}\label{eq:ODD:Price equation}
\begin{split}
\Delta \bar{z} = \underbrace{\mathrm{cov}(w, z)}_{\text{selection}} + \underbrace{\overline{w \Delta z}}_{\text{transmission}}
\end{split}
\end{align}

Where \(\bar{z}\) is the population mean trait value, \(\Delta \bar{z}\) its change over a time interval and \(w\) is relative fitness. The selection term is positive when trait values and fitness covary positively, driving \(\bar{z}\) upward, and is known as the selection differential (\(S\)). The transmission (\(T\)) term reflects systematic differences between the trait value of ancestors and their descendants, such as through mutation. \textcite{kline_designing_2018} propose analyzing group level cultural adaptations for resource use by decomposing the selection term of equation \eqref{eq:Price equation} into group and individual selection. We use the following formulation:

\begin{align}\label{eq:ODD:Selection differential}
\begin{split}
S =
\underbrace{\mathrm{cov}(w_g, z_g)}_{\text{group selection}} +
\underbrace{E\left[\mathrm{cov}(w_{ig}, z_{ig})\right]}_{\text{individual selection}}
\end{split}
\end{align}

Where \(z_g\) is the group mean trait value, \(w_g\) is the group mean relative fitness, \(w_{ig}\) is the individual relative fitness and \(z_{ig}\) is the individual trait value. \(E\) indicates a weighted average of the covariance among individuals within all groups, with the weights being the relative group size. Group means and among-group covariances are likewise weighted by group size, so that the group and individual components sum exactly to the population-level selection term.

We analyze the levels of selection using equation~\eqref{eq:ODD:Selection differential} and the MLS 1 method presented by \textcite{hermsen_emergent_2022}. Between time intervals $\Delta t$, the change in the population mean of a trait is tracked. Each agent is assigned a unique ID at $t_1$, becoming an ancestor, and at $t_2$ the number of descendants per ancestor is counted. If the individual was already born at time $t_1$, its past self is set as the ancestor. Descendants are agents who inherit traits of an ancestor, which can be biological (offspring) or cultural (imitation). We thus define term $S$ of equation~\eqref{eq:ODD:Selection differential} here as the \textit{cultural selection differential}. Absolute fitness ($W_i$) is defined as the number of descendants of agent (i) between time intervals, and relative fitness as ($w_i = W_i / \bar{W}$), with ($\bar{W}$) the population mean absolute fitness at that interval. We compute the components of the cultural selection differential at each time interval for behaviors towards harvesting, group boundary, resource boundary and monitoring.

The model includes mutation of behavioral strategies, contributing to the transmission term, which we do not analyze here as our interest is in the relative strength of selection at the two levels. The two components of \(S\) therefore only account for the selective part of \(\Delta \bar{z}\). In addition, note that there are numerous conflicting definitions of (cultural) fitness \cite{ramsey_is_2017}, and other definitions are possible. For example, cultural fitness might be considered relevant to wealth or mass. 

\subsection{Initialization}
The model is initialized a random amount of resources between 0 and $K$ assigned to each patch separately in the local treatment and or to the single resource in the global treatment. $N_0$ agents are spatially distributed in ten clustered groups of canonical institutional combinations randomly spread across the grid. Each cluster is assigned a random color from a fixed set of ten colors. After initialization, the processes described below are repeated every time-step. 

\subsection{Input data}
The model does not use input data to represent time-varying processes.

\subsection{Submodels}\label{sec:ODD:submodels}
\subsubsection{Territory defence}\label{par:ODD:group boundaries}
\begin{figure}[H]
    \centering
    \includegraphics[width=90mm]{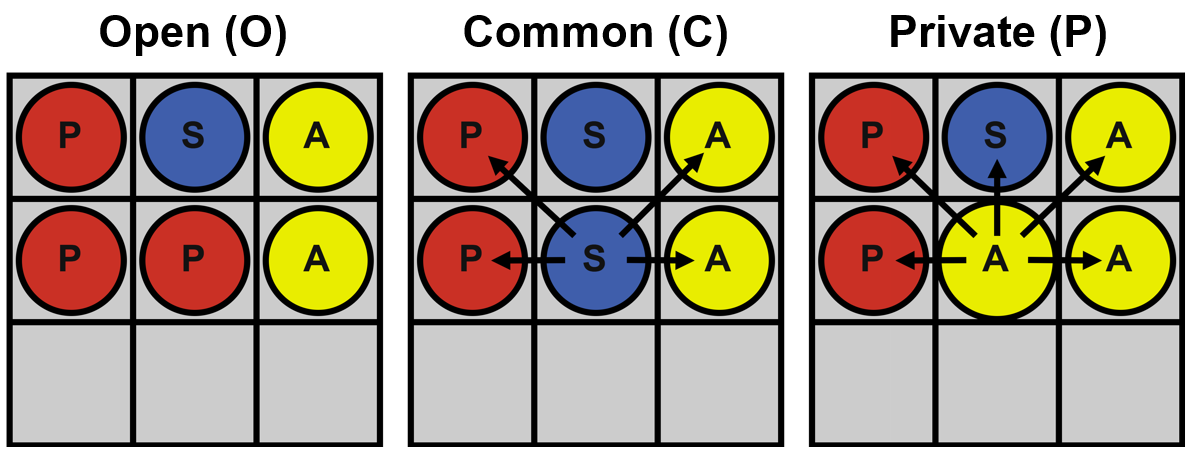}
    \caption{Agents may take one of the three institutional strategies (O, C, P) towards group boundaries.  Agents with open boundaries (O), will not attack anyone. Agents with common (C) boundaries will only attack neighbors with different social markers. Agents with private (P) boundaries will attack any neighbor. }
    \label{fig:ODD:Group boundaries}
\end{figure}

At the start of each timestep, group boundaries are operationalized through territory defense. A focal agent $i$ evaluates its Moore neighborhood for possible intruders and attacks according to their group boundary trait. Agents may hold one of three strategies towards boundary defense. Agents with open boundaries, will not attack anyone. Agents with common boundaries will only attack neighbors with different social markers. Agents with private boundaries will attack any neighbor (see figure \ref{fig:ODD:Group boundaries}). Each attacker/defender dyad is resolved at most once per time step. The outcome of the conflict is determined by the mass of the two parties in conflict with a single probabilistic draw:

\begin{equation}\label{eq:ODD:Conflict draw}
p_{\text{def}} =
\frac{M_{j,t, \text{def}}}
     {M_{j,t, \text{def}}
     + M_{j,t, \text{att}}}
\end{equation}

The losing party loses mass with attrition rate $\kappa$, and if this causes death, the winning agent can move to or reproduce on the empty patch left by the losing agent. In addition to possible death, both defending and attacking agents incur the cost of conflict due to an increased metabolic rate \ref{par:ODD:Allocation}. 

\subsubsection{Harvest}\label{par:ODD:Harvesting}
\begin{figure}[H]
    \centering
    \includegraphics[width=90mm]{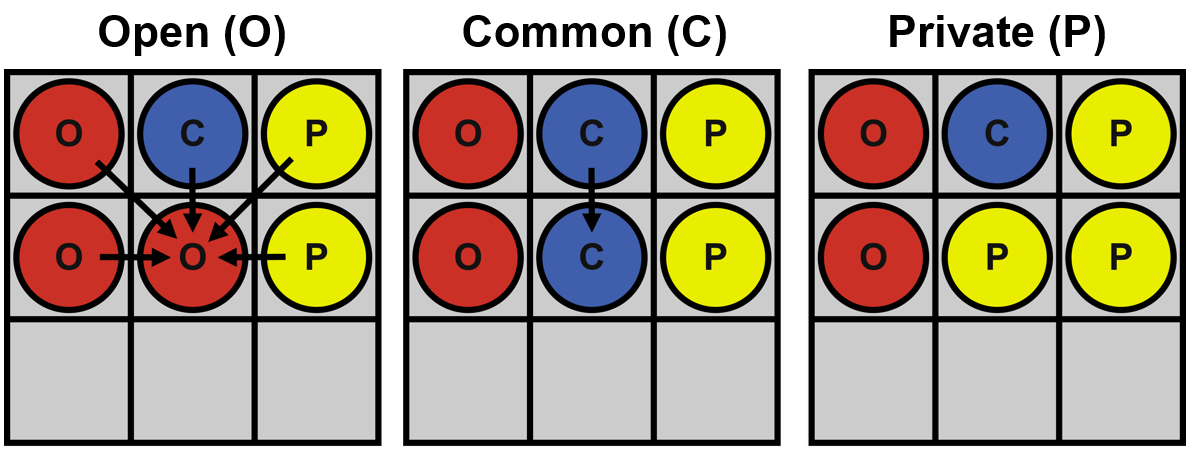}
    \caption{Agents may take one of the three institutional strategies (O, C, P) towards resource boundaries. Agents with an open access strategy (O) will allow shared access to their patch with anyone. Agents with a common access strategy (C) allow only agents that are of the same group marker. Agents with a private access strategy (P) will not allow anyone.}
    \label{fig:ODD:Resource boundaries}
\end{figure}

\begin{figure}[H]
    \centering
    \includegraphics[width=60mm]{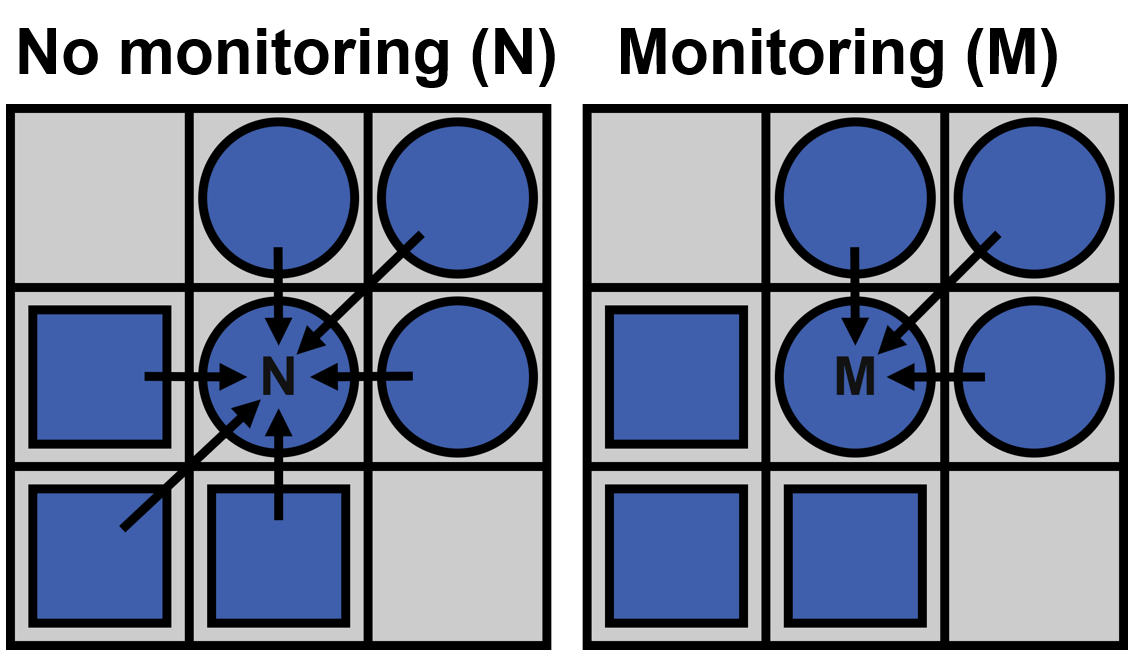}
    \caption{Agents may take one of the two institutional strategies (N, M) towards monitoring. Non-monitoring agents will allow shared access to their patch with anyone. Monitoring agents allow only agents that share the same harvesting preference. Agent (blue) shape represents harvesting preference (square = unrestrained, circle = restrained)}
    \label{fig:ODD:Monitoring}
\end{figure}

The agents then harvest from either local resources or a single global resource depending on the experiment treatment. In the local resource treatment, agents may share access to their resources with neighboring agents depending on their resource boundary strategy. Agents with open resource access will allow access to all other agents, commons agents will allow access to agents who share the same group marker, and private agents will not allow access to anyone (see figure \ref{fig:ODD:Resource boundaries}). In addition, depending on their monitoring trait, agents can monitor the harvesting strategy of agents whom they share access to, and consequently exclude or include them from shared harvesting (see figure \ref{fig:ODD:Monitoring}).

Harvesting rates depend on both the consumer and the resource variables. The maximum harvest rate ($I_{i, \text{max}}$) of a consumer scales allometrically with its mass ($M_{i}$): 

\begin{equation}\label{eq:Harvest scaling}
I_{i, \text{max}} = b_0 \, M_i^{3/4}
\end{equation}

Here, $b$ is a normalization constant. The harvest rate ($I_i$) then follows a saturating Holling type II functional response: 

\begin{equation}\label{eq:Harvest rate}
I_{i} = I_{\text{max}, i} \frac{E_{j}}{E_{j} + h}
\end{equation}

where $E_j$ denotes the available resource energy and $h$ is the half-saturation constant, i.e., the resource availability at which half of the maximum harvest rate is achieved. In the local resource setting, $E_j$ represents the total resource energy available across the agent’s local commons, defined by the cooperating set of neighboring agents. In the global resource setting, all agents harvest from a single shared resource pool $E_j$. The harvested resources are converted into resources available for allocation with assimilation efficiency $\epsilon$.

\subsubsection{Migration}\label{par:ODD:Migration}
Agents will migrate to empty neighboring patches if the potential gain is greater than the cost. The potential gain is the amount of resources on the neighboring patch with the most resources, minus the amount of resources on the agents' focal patch. Migration causes an increased metabolic rate, the cost is therefore the increased metabolic rate due to migration.

\subsubsection{Allocation}\label{par:ODD:Allocation}
Available resources at every timestep are the sum of the resources harvested in that step and the previously stored resources. At each time step, agents first incur a basal metabolic (BMR) maintenance cost that scales allometrically with body mass. Additional energetic costs are incurred when agents engage in energetically demanding activities such as conflict or migration. These activity-specific costs are modeled as multiplicative increases in metabolic expenditure relative to BMR. The resulting metabolic rate therefore represents the metabolic rate (FMR) of the agent during that timestep:

\begin{equation}\label{eq:ODD:FMR}
\mathrm{FMR}_{i, t}
= b_{\text{BMR}}\, M_i^{3/4}
\, \phi_i^{\text{fight}}
\, \phi_i^{\text{move}},
\end{equation}

where $b_{\mathrm{BMR}}$ is a normalization constant and  $\phi_i^{\text{fight}}$ and $\phi_i^{\text{move}}$ are behavioral cost multipliers. The available energy is reduced with $MR_{i,t}$. If the stored resources are below zero after this reduction, an agent is dead. 

After paying the energetic cost of maintenance and activities, any remaining energy is allocated to somatic growth. The energy of the assimilated resource restricts the rate at which the biomass consumed can be converted to consumer body mass, with the conversion efficiency $\epsilon_g$ determining losses during synthesis. Growth is also constrained by mass, agents follow a von~Bertalanffy-type growth formulation, such that the increase in body mass for agent $i$ at time $t$ is:

\begin{equation}\label{eq:ODD:Consumer growth}
M_{i,t+1} = M_{i,t} + \tau \left( M_{\max,i}^{1/3} M_{i,t}^{2/3} - M_{i,t} \right),
\end{equation}

where $M_{i,t}$ is the body mass of agent $i$ at time $t$, $M_{\max,i}$ is the agent-specific maximum body mass, and $\tau$ is a growth time constant.

Those with sufficient energy size and above a minimum size may reproduce if there is space, creating offspring with identical traits, except for mutations. Harvesting-, sharing- and conflict strategies together with social markers may mutate separately with probability $\mu$. Mass and energy passed on to offspring is removed from the parent.

\subsubsection{Imitation}\label{par:ODD:Imitation}
At each time step, agents compare their total mass ($M_i$), to that of neighboring agents. If a neighbor has a higher mass, the focal agent copies the traits of that neighbor with probability $\beta$.

\subsubsection{Age}\label{par:ODD:Age}
Agents have a age that increases by one unit each time step. Mortality from senescence is stochastic and increases logistically with age. At each time step, an agent with age $A_i$ dies with probability:

\begin{equation}\label{eq:ODD:Age}
P_{\text{death}}(A_i) = \frac{1}{1 + e ^ {10 - \frac{A_i}{8}}}
\end{equation}

Dead consumer mass and associated reserves are removed from the modeled system and treated as losses to an external sink.

\subsubsection{Resource regrowth}
Resource stocks regenerate each time step following density-dependent (logistic) growth:

\begin{equation}\label{eq:ODD:Biomass increase}
M_{j(t+1)} = M_{j(t)} + r M_{j(t)} \left(1 - \frac{M_{j(t)}}{K} \right)
\end{equation}

Where $K$ is the resource carrying capacity and $r$ is the intrinsic rate of biomass increase. In the local treatment, each patch follows Eq \ref{eq:ODD:Biomass increase} independently. In the global resource treatment a single global resource energy stock follows Eq \ref{eq:ODD:Biomass increase}. $K$ is calibrated so that the system-level carrying capacity is equal in both settings.

\subsubsection{Resource diffusion}
Patch resources diffuse locally at rate $D$, such that each patch transfers a fraction $D$ of its resources per time step equally among its eight neighbors; diffusion is disabled in the global resource setting.

\end{document}